\documentstyle[11pt,newpasp,epsf,twoside]{article}
\def\edcomment#1{\iffalse\marginpar{\raggedright\sl#1\/}\else\relax\fi}
\reversemarginpar

\def\kms{\relax \ifmmode {\,\rm km\,s}^{-1}\else \,km\,s$^{-1}$\fi}
\def\ha{\relax \ifmmode {\rm H}\alpha\else H$\alpha$\fi}
\def\hb{\relax \ifmmode {\rm H}\beta\else H$\beta$\fi}
\def\hi{\relax \ifmmode {\rm H\,{\sc i}}\else H\,{\sc i}\fi}
\def\hii{\relax \ifmmode {\rm H\,{\sc ii}}\else H\,{\sc ii}\fi}
\def\h2{\relax \ifmmode {\rm H}_2\else H$_2$\fi}

\def\fdg{\hbox{$.\!\!^\circ$}}

\def\farcm{\hbox{$.\mkern-4mu^\prime$}}
\def\farcs{\hbox{$.\!\!^{\prime\prime}$}}
\def\degd#1.#2{ #1\fdg#2 }                 
\def\mind#1.#2{ #1\farcm#2 }               
\def\secd#1.#2{ #1\farcs#2 }               

\begin{document}
\title{From gas to starburst and AGN}
\author{Johan H. Knapen}
\affil{Department of Physics, Astronomy and Mathematics, University of
Hertfordshire, Hatfield, Herts AL10 9AB, UK}

\begin{abstract}

Although it is clear that the circumnuclear regions of galaxies are
intimately related to their host galaxies, most directly through their
bars, it remains unclear what exactly initiates and fuels nuclear
stellar (starburst) and non-stellar (AGN) activity. Deviations from
axisymmetry in the gravitational potential of a galaxy, set up by a
bar or an interaction, are known to cause gas inflow, and must at some
scale and level be related to the fueling of AGN and starbursts. We
review the observed relations between bars and interactions on the
one, and nuclear activity of the Seyfert and starburst variety on the
other hand, and conclude that none of these relations is particularly
significant in a statistical sense, except in extreme and rare cases,
such as ultra-luminous infrared galaxies. Nuclear rings of star
formation, however, are not only related directly to non-axisymmetries
in the potential, but also, it seems, to the occurrence of nuclear
activity. Their role as potential tracers of the fueling process must
be further explored.

\end{abstract}

\section{Introduction}

Nuclear activity in galaxies, either of the stellar (a nuclear or
circumnuclear starburst) or non-stellar (``active galactic nucleus'';
AGN) variety, is rather common in present-day galaxies (e.g., Ho,
Filippenko, \& Sargent 1997a).  The most tangible link between the
nuclear regions and the host galaxy is provided by the fact that the
mass of the central supermassive black hole in a galaxy is directly
related to the velocity dispersion (hence the mass) of the bulge
(Ferrarese \& Merritt 2000; Gebhardt et al. 2000). Since such black
holes appear to be ubiquitous in the nuclei of both active and
non-active galaxies (Kormendy \& Richstone 1995), whether or not a
galaxy is active must largely be determined by the ability of that
galaxy to channel digestible fuel (i.e., gas) to the smallest scales.
Large stellar bars, as well as tidal interactions and mergers, can
drive gas efficiently from the outer disk into the inner kpc
(Shlosman, Frank, \& Begelman 1989; Shlosman, Begelman \& Frank 1990),
but it is non-trivial to drive gas to smaller scales.  Nuclear
(secondary) bars nested within the large-scale (primary) stellar bar
(e.g., Shlosman et al.  1989; Laine et al.  2002) have been suggested
as vehicles which drive gas efficiently to scales of 10--100 pc.

We will concentrate here on observational evidence, mostly statistical
in nature, of the effects of bars and interactions on starburst and
Seyfert activity, which are conceivably linked to these facilitators
of inflow. After reviewing observations which can support the
theoretical expectation that bars will concentrate mass in the center
of a galaxy (Sect.~2), we will review separately the observational
evidence, or lack thereof, for relationships between the forms of
non-axisymmetry and activity identified above (Sect.~3-6). We will
then discuss nuclear rings of star formation and their relations with
bars and nuclear activity (Sect.~7, 8). Concluding remarks are given
in Sect.~9.

\section{Bars and central mass concentration}

There are several pieces of observational evidence which indicate that
bars do indeed concentrate gas in the central regions of spiral
galaxies, as expected theoretically and numerically. We will briefly
review the evidence here, but also emphasize that it does not
necessarily imply that the mass concentration stimulated by bars is
connected directly to the occurrence or fueling of nuclear activity
because the scales involved are different. Such a connection between
mass concentration and activity would have to be separately studied in
a statistically meaningful way, which has to our knowledge not been
done to date.

The most comprehensive evidence so far for mass concentration by bars
stems from surveys of molecular gas concentration in barred as
compared to non-barred galaxies, traced through emission in the
millimeter domain by CO molecules. Sakamoto et al. (1999) used data
from the Owens Valley and Nobeyama interferometers for 10 barred and
10 non-barred galaxies, and found that the barred galaxies,
statistically, have more concentrated CO emission, and thus,
presumably, molecular hydrogen. More recently, Sheth et al. (2004)
used the data from the BIMA-SONG survey (Regan et al. 2001; Helfer et
al. 2003) to confirm this conclusion.  Their sample contains 29 barred
and 15 non-barred galaxies of types Sab-Sd, which have comparable
molecular gas masses and surface densities across their disks. In the
central kiloparsec, however, both the average surface density and the
central concentration of molecular gas are about three times higher in
the barred than in the non-barred galaxies, effects which are somewhat
more pronounced in the early than in the late type galaxies in the
sample. Although the galaxies with the most centrally concentrated CO
emission and the highest surface densities in the central kiloparsec
are barred, the scatter is large. One of the caveats in this kind of
work is the value of the conversion factor between CO intensity and
molecular hydrogen column density, the $X$-factor, which is usually
assumed constant. Findings such as the one by H\"uttemeister et
al. (2000) of significant changes in the conversion factor within the
barred galaxy NGC~7479 as determined from $^{12}$CO/$^{13}$CO line
intensity ratios, possibly related to the gas being more diffuse, or
unbound, in the bar, may not invalidate this general assumption, but
do highlight the need for more detailed scrutiny.

Maiolino, Risaliti, \& Salvati (1999) studied the atomic gas column
density in front of Seyfert nuclei using X-ray measurements, and found
significantly higher columns in those Seyferts which are in barred
hosts than in those in non-barred galaxies. Alonso-Herrero \& Knapen
(2001) studied \hii\ region populations in the central regions of
spiral galaxies, and found that both the luminosity of the most
luminous \hii\ regions and the massive star formation efficiency are
higher in the barred than in the non-barred galaxies, which again was
interpreted as a consequence of enhanced gas concentration by
bars. There is thus indeed some observational evidence that bars drive
gas towards the nuclear zones of galaxies. Below, we will explore
whether this can be related to the occurrence of nuclear starburst or
Seyfert activity.

\section{Bars and starburst activity}

Clear indications that nuclear starbursts preferentially occur in
barred hosts have been reported since the early eighties, from radio
continuum (e.g., Hummel 1981; Puxley, Hawarden, \& Mountain 1988;
Huang et al. 1996) or infrared (e.g., Hawarden et al. 1986; Devereux
1987; Dressel 1988) emission and its distribution, or from the
morphology of spectroscopically selected starburst galaxies (Arsenault
1989). Even though the results are not unambiguous (see, e.g.,
discussions in Huang et al. 1996), there is a clear trend. For
example, Hummel (1981) found that the central radio continuum
component is typically twice as strong in barred than in non-barred
galaxies; Hawarden et al. (1986) found that basically all galaxies
with a high 25/12$\mu$m flux ratio are barred; and Arsenault (1989)
found an enhanced bar$+$ring fraction among starburst hosts. Huang et
al. (1996) refined the methodology used in earlier papers and used
IRAS data to confirm that starburst hosts are preferentially
barred. They do point out, however, that this result only holds for
strong bars (SB class in the RC3 catalog, de Vaucouleurs et al. 1991)
and in early-type galaxies. All results mentioned above are based on
morphological classifications taken from catalogs such as the RC3.

A study of the molecular gas properties in the inner kpc of starbursts
and non-starbursts (Jogee 2001; Jogee et al. 2001) shows that the
starbursts generally host larger central gas densities (assuming a
standard $X$-factor).  The fact that both starbursts and
non-starbursts in Jogee's sample can host a large-scale bar suggests
that the lifetime of the starburst is short with respect to the
timescale over which bars evolve or dissolve. The data also suggest
that in a given barred galaxy the dominant star formation mode can
change from an inefficient pre-burst phase in the early stages into a
powerful circumnuclear starburst once a high enough central gas
density builds up.  These findings are consistent with the reports
referred to above that starbursts are more common in barred than in
unbarred galaxies, but they also imply that the presence of a bar is
not a sufficient condition for a concurrent starburst.

We conclude that statistical studies show that bars and starbursts are
connected, but that the results are subject to important caveats and
exclusions. Further study is needed, using carefully defined samples,
and exploring more direct starburst indicators, such as nuclear
spectra, H$\alpha$ imaging, or possibly careful SED fitting, all
combined with a bar analysis based upon, ideally, near-infrared (NIR)
imaging. What also needs more scrutiny is the nature of the starburst
associated with a bar. Hawarden et al. (1986) already suggested that
these starbursts might be circumnuclear, and we now know that such
circumnuclear star formation can be very compact (e.g., Gonz\'alez
Delgado et al. 1998). Given the theoretically {\it and}
observationally established strong connections between circumnuclear
star formation and non-axisymmetry in the gravitational potential of
their host galaxy (Sect.~7), and given the importance of such features
for our understanding of galaxy dynamics and evolution, {\it HST}
imaging ought to be employed to investigate whether bar-related
starburst activity is circumnuclear rather than nuclear in general.

\section{Bars and Seyfert activity}

Seyfert galaxies have a number of characteristics which make them
particularly suitable for a study of their host galaxies: they are
relatively local AGN and occur predominantly in disk galaxies.  Over
the years, and starting with the work of Adams (1977), a fair number
of authors have studied the fraction of bars in Seyfert galaxies,
often comparing the results for the AGN with those for a control
sample (e.g., Adams 1977; Simkin, Su, \& Schwarz 1980; Balick \&
Heckman 1982; MacKenty 1990). Unfortunately, these early surveys, and
many of the later ones (e.g., Moles, M\'arquez, \& P\'erez 1995; Ho,
Filippenko \& Sargent 1997b; Crenshaw, Kraemer, \& Gabel 2003) suffer
from one or more of the following imperfections: they are based on
optical imaging, where stellar bars are much more difficult to detect
than in the NIR; they use the RC3 classification or, worse, ad-hoc and
non-reproducible classification criteria to determine whether a galaxy
is barred; or they suffer from the absence of properly matched control
samples.

There have also been a small number of studies using new,
high-quality, NIR imaging of well-matched samples of Seyfert and
quiescent galaxies. Mulchaey \& Regan (1997), from one such study,
report identical bar fractions, but Knapen, Shlosman, \& Peletier
(2000), using imaging at higher spatial resolution and a rigorously
applied set of bar criteria, find a marginally significant difference,
with a higher bar fraction in a sample of CfA Seyferts than in a
control sample of non-Seyferts (approximately 80\% vs. 60\%).

This difference was later confirmed at a 2.5 $\sigma$ level by Laine
et al. (2002), who improved upon the work by Knapen et al. (2000) by
increasing the sample size (112 instead of 58 galaxies in total) and
by using high resolution {\it HST} NICMOS NIR images of the central
regions of all galaxies. Laine et al. (2002) find that 41 of their 56
Seyfert galaxies have at least one bar (73\%$\pm$6\%, where the
uncertainty is the Poisson error due primarily to the sample size),
against 28 of the 56 non-Seyfert control galaxies
(50\%$\pm$7\%). Another result is that almost one of every five sample
galaxies, and almost one of every three barred galaxies, have more
than one bar, although the fraction of small, or nuclear, bars is not
enhanced in Seyfert galaxies (see also Erwin \& Sparke 2002, who reach
much the same conclusions on nuclear bars, even though the
classification of especially the smallest bars remains a matter of
debate for some individual galaxies, see, e.g., Laine et al. 2002;
Erwin 2004). Laine et al. also find that there is no correlation
between the presence of companion galaxies, even relatively bright
ones, and a bar, thus refuting a claim by M\'arquez et al. (2000) that
the results by Knapen et al. (2000) were unreliable because not all
the sample galaxies were isolated.

We can thus conclude that carefully performed studies show that there
is a slight, though only just about significant, excess of bars among
Seyfert galaxies as compared to non-Seyfert control samples. A
significant fraction of Seyfert galaxies appear to be non-barred,
although future analysis could potentially reveal a weak or very small
bar, or another form of non-axisymmetry in their gravitational
potential which could have similar dynamical effects (Shlosman et
al. 1989). What even the most detailed of these studies have not
shown, however, is a one-to-one correlation between bars and Seyfert
activity. Given that any fueling process must be accompanied by
angular momentum loss, which in turn is most likely induced by
non-axisymmetries, either the timescales of bars (or interactions, see
below) are different from those of the activity, or the
non-axisymmetries are not as easy to measure as we think, for instance
because they occur at smaller scales, or are masqueraded to a
significant extent by, e.g., dust or star formation (Laine et
al. 2002).

\section{Interactions and starburst activity}

There is ample anecdotal evidence for the connection between galaxy
interactions and starburst activity. Practically all extreme infrared
sources, specifically the Ultra-Luminous InfraRed Galaxies (ULIRGs),
which are thought to be powered mainly by extreme starbursts (Genzel
et al. 1998), occur in galaxies with disturbed morphologies,
presumably interacting (e.g., Joseph \& Wright 1985; Armus, Heckman,
\& Miley 1987; Sanders et al. 1988; Clements et al. 1996; Murphy et
al. 1996; Sanders \& Mirabel 1996; see also Sanders, these
proceedings, p.~000). Partly as a result of this finding, and
considering the much increased interaction rates at large lookback
times expected from cosmological models, a significant occurrence of
massive starbursts, mostly dust-obscured and thus hidden from direct
view, is sometimes inferred from the observation of so-called
SCUBA-sources (e.g., Smail, Ivison, \& Blain 1997). Given the evidence
from, for instance, observations of ULIRGs, it is perhaps not
unreasonable to state that such massive starbursts are powered in
galaxies which are undergoing a major upheaval, i.e., are merging or
interacting.

More in general though, the picture is far less clear, as nicely
illustrated in a recent paper by Bergvall, Laurikainen, \& Aalto
(2003). These authors considered two matched samples of nearby
interacting (pairs and clear cases of mergers) and non-interacting
galaxies, and measured star formation indices based on $UBV$
colors. Contrary to previous reports in the literature (notably by
Larson \& Tinsley 1978), Bergvall et al. do not find significantly
enhanced star-forming activity among the interacting/merging galaxies.
Bergvall et al. (2003) estimate from their sample that only about
0.1\% of a magnitude limited sample of galaxies will be massive
starbursts generated by interactions and mergers. The authors then
argue that since virtually all ULIRG morphologies show evidence for
recent mergers, other mechanisms cannot reasonably be expected to
trigger massive starbursts. Although Bergvall et al.'s fraction of
0.1\% is based upon one true massive starburst (ESO\,286-IG19) found
in a sample of 59 and thus carries substantial error bars, it is also
clear that such massive starbursts are a very rare occurrence indeed,
at least, but possibly not exclusively (Bergvall et al.), in the local
universe.

So although mergers can undoubtedly lead to massive starbursts, and
most extreme starbursts show evidence for a merging/interacting
history, they appear to do so only in exceptionally rare cases. Most
interactions between galaxies may not lead to any increase in the
starburst activity, and those that do may be selected cases where a
set of parameters, both internal to the galaxies and regarding the
orbital geometry of the merger, is conducive to the occurrence of
starburst activity (e.g., Mihos \& Hernquist 1996).  Laine et
al. (2003) find very little evidence for trends in starburst activity
from detailed {\it HST} imaging of the Toomre sequence of merging
galaxies, further illustrating this point.

\section{Interactions and Seyfert activity}

Galaxy interactions can easily lead to non-axisymmetries in the
gravitational potential of one or more of the galaxies involved, and
as such can be implicated in angular momentum loss of inflowing
material, and thus conceivably in AGN fueling (Shlosman et al. 1989).
Although indeed Seyfert activity occurs in interacting and merging
galaxies (NGC~2992 is a rather spectacular example), no unambiguous
evidence has been reported for an excess of the numbers of companions
to Seyfert galaxies as compared to non-active control galaxies (e.g.,
Fuentes-Williams \& Stocke 1988; de Robertis, Yee, \& Hayhoe 1998),
nor for a different incidence of Seyfert or AGN activity among more or
less crowded environments (e.g., Kelm, Focardi, \& Palumbo 1998;
Miller et al. 2003). Much of the earlier work, some of which claimed a
statistical connection between interactions and Seyfert activity, was
unfortunately plagued by poor control sample selection (see
Laurikainen \& Salo 1995 for a detailed review), while most studies,
including some of the most recent ones, are not based on complete sets
of redshift information for the possible companion galaxies.  In
addition, Laine et al. (2002) have shown that the bar fraction among
both their Seyfert and non-Seyfert sample galaxies is completely
independent of the presence of faint, or bright, companions
(interacting galaxies were not considered by Laine et al.).  We thus
conclude that interactions and Seyfert activity may well be linked in
individual cases, but that as yet the case that they are statistically
connected has not been made convincingly.

\section{Bars and nuclear rings}

\begin{figure}
\plotone{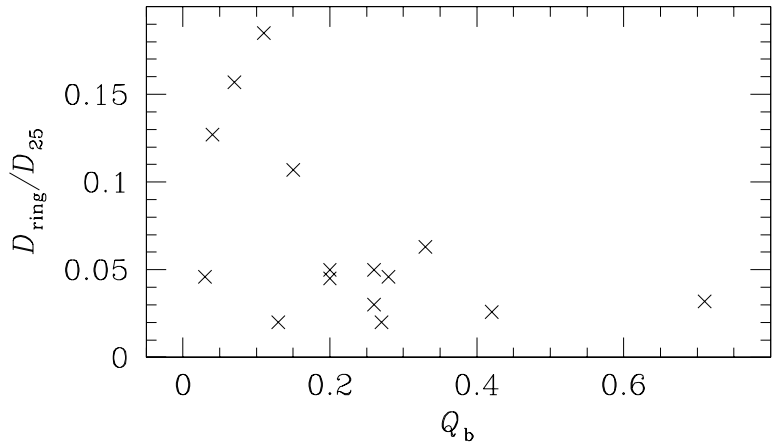}
\caption{Relative size (ring diameter
divided by host galaxy diameter) for a sample of 15 nuclear rings as a
function of the gravitational torque $Q_{\rm b}$, or strength, of the
bar of its host galaxy. Data from Knapen, P{\' e}rez-Ram{\'{\i}}rez, \&
Laine (2002) and Knapen (2004).}
\label{ringdiam}
\end{figure}

As discussed earlier in this paper, bars concentrate gas in the
central regions of galaxies. Bars, however, also set up resonances
which can act as focal points for the gas flow, and where gas
concentrates in limited radial ranges (see, e.g., review by Shlosman
1999). Rings in disk galaxies are intimately linked to the internal
dynamics and the evolution of their hosts, are relatively common, and
are most often outlined by massive star formation (see Buta \& Combes
1996 for a comprehensive review on galactic rings).

Nuclear rings, on scales of less than one to roughly two kiloparsec in
radius, occur in one of every five spiral galaxies (Knapen 2004).
They are found almost exclusively in barred galaxies (e.g., Buta \&
Combes 1996; Knapen 2004; see below), and can be directly linked to
inner Lindblad resonances (Knapen et al. 1995; Heller \& Shlosman
1996; Shlosman 1999). Individual gas clouds within such a ring can
become gravitationally unstable, either spontaneously (Elmegreen
1994), or under the influence of density waves set up by the bar
(Knapen et al. 1995; Ryder, Knapen, \& Takamiya 2001), and rings are
well known for the significant massive star formation occurring within
them (although rings without star formation occur as well, see, e.g.,
Shlosman 1999; Erwin \& Sparke 2002). Because of the much enhanced
massive star formation occurring in nuclear rings, they are prime
tracers of star formation processes in starburst regions, as well as
of the dynamics of galaxies on scales of a kiloparsec from the
nucleus. As an example of the latter, Fig.~\ref{ringdiam} shows the
relation between the relative nuclear ring size and the gravitational
bar torque (a measure of the strength of the bar) for 15 nuclear
rings. It is clear that large rings can only occur in weak bars, a
result which confirms expectations from theory and modeling, in which,
simply put, the extent of the perpendicular $x$2 orbits needed to
sustain the nuclear ring is limited as the bar gets stronger, i.e., as
the $x$1 orbits become more elongated (see Knapen et al. 1995; Heller
\& Shlosman 1996; Knapen, P\'erez-Ram\'\i rez, \& Laine 2002; Knapen
2004).

Some rings apparently occur in non-barred galaxies. In some cases, the
host may be classified as non-barred in the major catalogs, but a
small bar shows up clearly in NIR imaging (for instance, NGC~1068,
Scoville et al. 1988, and NGC~4725, Shaw et al.  1993; M\"ollenhoff,
Matthias, \& Gerhard 1995; see the more elaborate discussion in Knapen
2004). Other apparently non-barred nuclear ring hosts may either have
a weak oval distortion, or be undergoing the effects of an interaction
with a companion galaxy. In either case, a departure from axisymmetry
would be set up in the gravitational potential of the galaxy which
could lead to ring formation in a very similar way to when a bar is
present (Shlosman et al. 1989). We give two examples from the
recent literature.

\begin{figure}
\plotfiddle{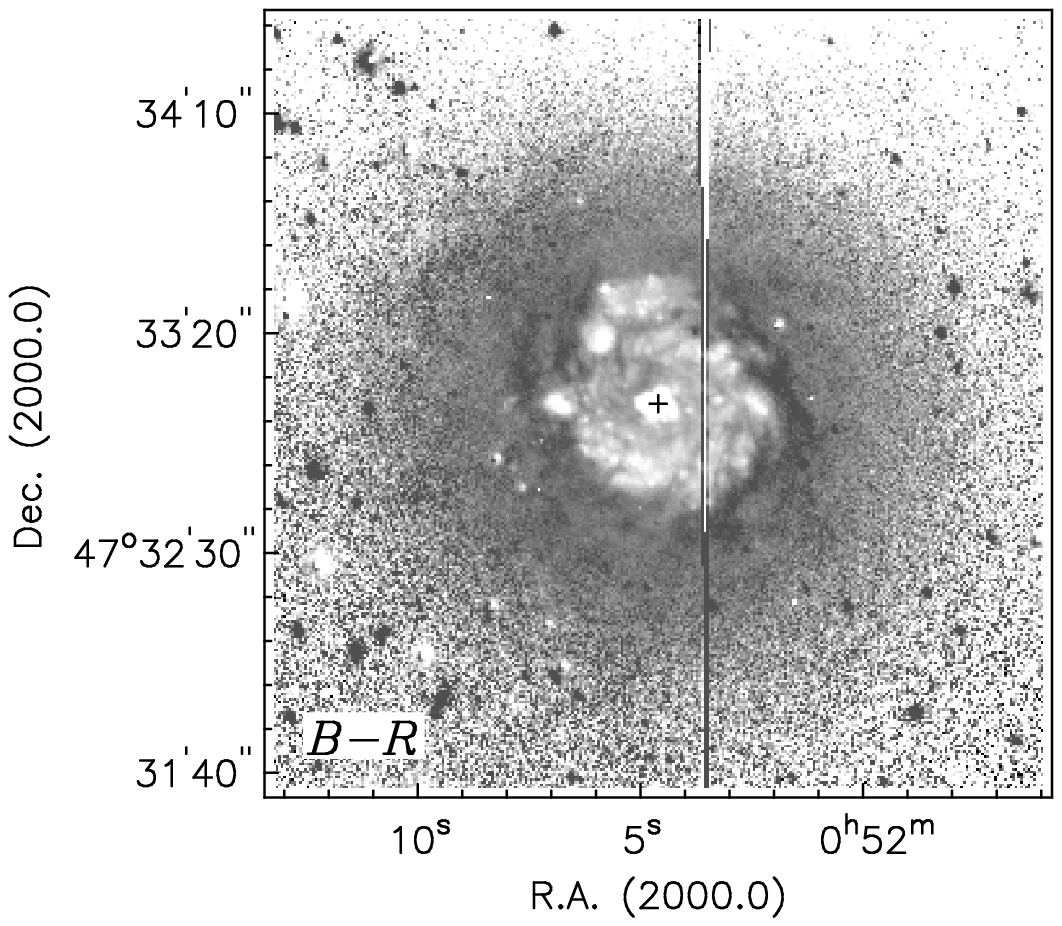}{4.8cm}{0}{55}{55}{-210}{-160}
\plotfiddle{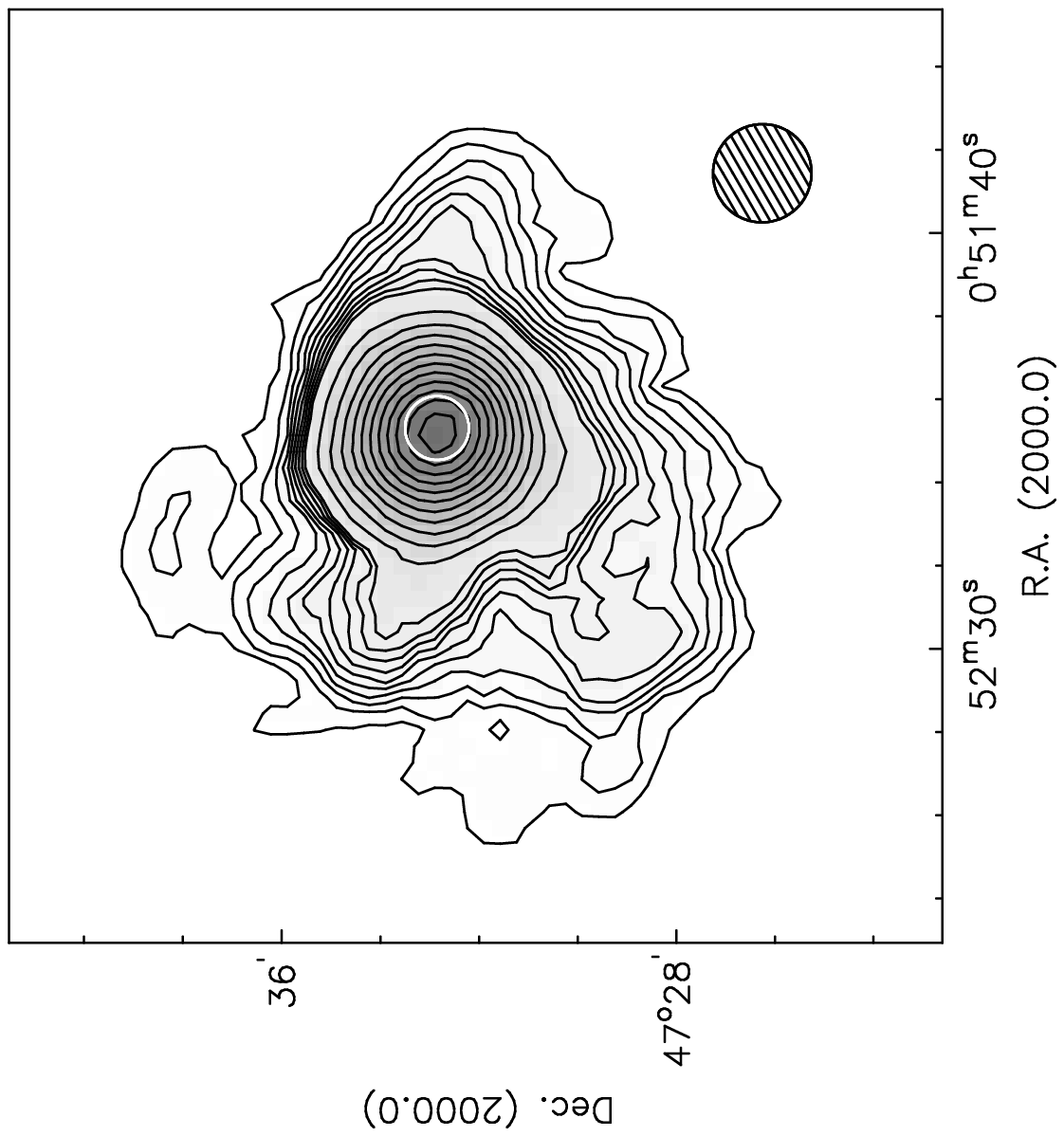}{0cm}{-90}{44}{44}{-30}{267}
\vspace{-1cm}
\caption{$B-R$ ({\it left}) and \hi\ ({\it right}) views of the galaxy
NGC~278. The scales covered in the two panels are very different: the
total extent of the optical disk of the galaxy, more or less twice the
area shown on the {\it left}, is indicated by the small white circle
in the middle of the {\it right} panel. The $B-R$ color index image
shows bluer colors as lighter shades. The \hi\ image shows the
integrated \hi\ surface density distribution at 120~arcsec resolution,
with contour levels at $(1, 2, 3, ...  , 9)\times10^{19}, (1, 1.5, 2,
..., 6.5)\times10^{20}$\,cm$^{-2}$. The vertical spike in the $B-R$
image is an artifact. Data from Knapen et al. (2004a).}
\label{278fig}
\end{figure}

The first example is that of NGC~278, a small, nearby and isolated
spiral galaxy ($v_{\rm sys}=640$\,km\,s$^{-1}$; $M_B=-18.8$) recently
observed in optical and \hi\ by Knapen et al. (2004a). Although
classified as SAB(rs)b in the RC3, there is no evidence for the
presence of a bar in this galaxy from either {\it HST} WFPC2 or
ground-based NIR imaging.

As seen in Fig.~\ref{278fig} ({\it left}), a $B-R$ color index image of
NGC~278, its optical disk shows two distinct regions, the inner one
with copious star formation and clear spiral arm structure, and the
outer one ($r>27$~arcsec or 1.5~kpc) which is almost completely
featureless, of low surface brightness, and rather red. The \hi\ disk
(shown at a low resolution of 120~arcsec in Fig~\ref{278fig}, {\it
right}) is much more extended than the optical disk (indicated by the
white circle in the \hi\ figure) and shows both morphological
(Fig.~\ref{278fig}) and kinematic (Knapen et al. 2004a) disturbances
which suggest a recent minor merger with a small gas-rich galaxy,
perhaps similar to a Magellanic cloud.

The scale and morphology of the region of star formation in NGC~278
indicate that this is in fact a nuclear ring, albeit one with a much
larger {\it relative} size with respect to its host galaxy than
practically all other known nuclear rings (the absolute radius of the
nuclear ring is about a kiloparsec, normal for nuclear rings). Knapen
et al. (2004a) postulate that it is in fact the past interaction which
has set up a non-axisymmetry in the gravitational potential, which in
turn, in a way very similar to the action of a classic bar, leads to
the formation of the nuclear ring. The case of NGC~278 illustrates how
in apparently non-barred galaxies rings can be caused by
departures from axisymmetry induced by interactions, but also shows how
difficult it can be to uncover this: in the case of NGC~278 only
through detailed \hi\ observations.

Another example of the same class may be the case of the blue compact
dwarf galaxy Mrk~409, which has recently been shown to possess two
star-forming rings, a nuclear one at 0.5~kpc radius, and a further one
at a radius of some 2~kpc (Gil de Paz et al. 2003). These authors
interpret the inner of the two rings as resulting from a
starburst-driven shock interacting with the interstellar medium. In
such a scenario, the nuclear starburst would sweep up gas in the
galactic plane, which would eventually reach densities high enough for
it to become molecular, followed by a phase of massive star
formation. Such a scenario seems unattractive, by the way, as a
general one, because it would depend on a very strong starburst-bar
correlation to reproduce the observed very strong nuclear ring-bar
correlation (the former has not been established nearly as
conclusively as the latter, see Sect.~3). Gil de Paz et al. (2003)
do not offer a preferred explanation for the outermost of the two
observed rings in this galaxy.

In light of the appearance and scales of the two rings in Mrk~409, it
is tempting to consider them as resonance rings. In that case, as in
host galaxies of other nuclear rings, we would expect to see some
evidence for the presence of a non-axisymmetry in the gravitational
potential. Also known as NGC~3011, Mrk~409 has been classified as
`.L.....' in the RC3. No {\it HST} images are available from the
archive, and a 2MASS (Jarrett et al. 2003) image shows no evidence for
a bar.  We checked the Hyperleda database (Paturel et al. 2003), and
found a number of companion galaxies which, although probably not
directly interacting, are certainly close enough to be, or to have
been, in some kind of gravitational partnership with Mrk~409. These
include PGC~028169 with $v_{\rm sys}=1557$\,km\,s$^{-1}$ and at a
distance of less than 250~kpc, UGC~5287 ($v_{\rm
sys}=1470$\,km\,s$^{-1}$; $250<d<300$~kpc), and UGC~5282 ($v_{\rm
sys}=1470$\,km\,s$^{-1}$; $300<d<350$~kpc). All these galaxies have
similar magnitudes to Mrk~409, which has a $v_{\rm sys}$ of
1527\,km\,s$^{-1}$. So there are clearly companion galaxies around
which may have caused non-axisymmetry in the potential, possibly as in
NGC~278. Unfortunately, \hi\ data are not available in this case, but
we speculate that these might well show disturbed kinematics and
morphology in the outer regions.

\section{Nuclear rings and nuclear activity}

Although nuclear rings and especially nuclear activity have as
separate topics received considerable attention in the literature,
their possible interrelation has not been much studied. Many nuclear
rings, of course, are anecdotally known to occur in galaxies which
also host a nuclear starburst or a prominent AGN (often of Seyfert or
LINER type, given the typical parameters of the host galaxies
involved), and some famous examples include NGC~1068 and NGC~4303.

In a recent paper, we explored the correlations between nuclear
activity (both of the non-stellar and starburst variety) and the
occurrence of nuclear rings in a sample of 57 nearby spiral galaxies
(Knapen 2004). Using information on the activity from the NASA/IPAC
Extragalactic Database (NED) and ring parameters from our own
H$\alpha$ imaging survey (Knapen et al. 2004b), we found not only that
nuclear rings significantly more often than not occur in galaxies
which also host nuclear activity (only two of the 12 nuclear rings
occur in a galaxy which is neither a starburst nor an AGN host; 30 of
the 57 sample galaxies would fall into this category), but also that
the circumnuclear H$\alpha$ emission morphology of the AGN and
starbursts is significantly more often in the form of a ring than in
non-AGN, non-starburst galaxies (38\% of AGN, 33\% of starbursts, 11\%
of non-AGN, and 7\% of non-AGN non-starburst galaxies have
circumnuclear rings in our sample of galaxies).

Although the number of galaxies in this initial study is rather small
for detailed statistical analyses, we did find this most interesting
correlation between the occurrence of nuclear rings and that of
nuclear activity. Our initial interpretation of this effect is that
both nuclear rings, as traced by the massive star formation within
them, and starbursts and AGN (of the Seyfert or LINER variety) trace
very recent gas inflow.  Although it is not clear {\it a priori} why
the kpc-scale fueling of nuclear rings and the pc-scale fueling of
activity might be so closely related, nuclear rings do seem to show a
potential for being very interesting and direct tracers of AGN
fueling. These findings may also be related to the reported higher
incidence of rings (inner and outer, as based on RC3 classifications)
among Seyfert and LINER hosts than among normal or starburst galaxies
(Hunt \& Malkan 1999). All these aspects of rings and nuclear activity
need further scrutiny.

\section{Concluding remarks}

There is significant evidence that bars can drive gaseous material to
the central regions of a galaxy (Sect.~2), but it is hard to pin down
the evidence that this gas leads directly to AGN or even starburst
activity. There are indications that starbursts are provoked by bars
or interactions, though most probably not in general, but only in
special cases (Sect.~3, 5). For Seyfert galaxies, there is very little
evidence indeed that this kind of AGN preferentially occurs in host
galaxies which are either barred (Sect.~4) or interacting
(Sect.~6). These rather disappointing observational results, however,
do not imply that bars and interactions can be declared altogether
innocent of being involved with nuclear activity.

The kind of deviations from axisymmetry in the gravitational potential
of the host galaxy set up by bars and interactions has, theoretically
and numerically, long since been linked to nuclear activity (e.g.,
Shlosman et al. 1989, 1990; see also review by Shlosman
2003). Shlosman et al. postulated that the presence of such
non-axisymmetry would be a necessary, but not sufficient condition for
the onset of nuclear activity, and that an additional factor or
factors, such as the availability of gas at the relevant (small)
scales, would likely play a role.

Observationally, we now know that bars can lead to central gas
concentration, thus assuring that a viable gas reservoir is present
within the central kiloparsec. Either many (or most, or possibly even
all) galaxies then become a starburst or Seyfert-type AGN at some
stage, but for a short time (and possibly eradicating any links we may
wish to find with the host's properties; Beckman 2001), or some
mechanism, as yet unidentified observationally, turns on the activity
in specific galaxies. This mechanism may well include gravitational
torques due to a non-axisymmetric matter distribution, for instance
due to nuclear bars (e.g., Laine et al. 2002), which can dump fuel
very close to the nucleus indeed (Shlosman et al. 1989). The
timescales of such nuclear inflow are unknown, but may be very short.

Nuclear bars are, unfortunately, rather hard to study
observationally. Morphological studies based upon, e.g., {\it HST} NIR
imaging are hampered by the effects of dust extinction and star
formation (Laine et al. 2002). The morphology of dust lanes in the
central kpc has been connected to attempts to identify nuclear bars
(Regan \& Mulchaey 1999; Martini et al. 2001), but nuclear bars,
unlike large bars, cannot be expected to lead to a well-defined
morphology of offset dust lanes (Shlosman \& Heller 2002; Maciejewski
et al. 2002). A possible way forward here is the use of integral field
spectroscopy (e.g., Bacon et al. 2001), which, if used in conjunction
with adaptive optics techniques, gives simultaneous high-resolution
mapping of the distributions of stellar populations and dust, as well
as of the gas and stellar kinematics. In combination with detailed
numerical modeling, this could lead to the detection of the dynamical
effects of a nuclear bar on gas flows which may be directly related to
the fueling process of starbursts and/or AGN.

As indicated by the many, often contradictory, studies referenced in
this paper, there is, however, also still a dire need for continued
study of the detailed properties of large and well-defined samples of
AGN and starburst host galaxies, as well as of carefully matched
control samples of non-active galaxies. Tantalising new results,
briefly described in Sect.~7 and~8, indicate that nuclear rings ought
to be included in such investigations. The current or future
availability of datasets of unprecedented scope and size (e.g., the
2MASS or Sloan surveys) must be exploited to identify and analyse any
relationships between nuclear activity and the properties of their
host galaxies, which are no doubt more subtle than often suggested in
the past.

{\it Acknowledgments} 

I wish to thank Shardha Jogee, Seppo Laine, and Isaac Shlosman for
comments on an earlier version of this paper, and Erwin de Blok for
assistance with Fig.~\ref{278fig}. I am especially indebted to my
collaborators on the various aspects of our joint research described
here.

\end{document}